\documentclass[acmlarge,screen]{acmart}

\usepackage[ruled, vlined]{algorithm2e}

\AtBeginDocument{%
  \providecommand\BibTeX{{%
    \normalfont B\kern-0.5em{\scshape i\kern-0.25em b}\kern-0.8em\TeX}}}

\setcopyright{acmcopyright}
\copyrightyear{2023}
\acmYear{2023}
\acmDOI{XXXXXXX.XXXXXXX}

\acmConference[ICCMB '24]{Proceedings of the 2024 7th International Conference on Computers in Management and Business}{Jan 12--14, 2024}{Singapore}

\begin{document}

\title{A General Framework on Enhancing Portfolio Management with Reinforcement Learning}

\author{Yinheng Li}
\authornote{All authors contributed equally to this research.}
\email{yl4039@columbia.edu}
\affiliation{%
  \institution{Columbia University}
  \city{New York}
  \state{NY}
  \country{USA}
  \postcode{10027}
}

\author{Junhao Wang}
\authornotemark[1]
\email{jw3668@columbia.edu}
\affiliation{%
  \institution{Columbia University}
  \city{New York}
  \state{NY}
  \country{USA}
  \postcode{10027}
}

\author{Yijie Cao}
\authornotemark[1]
\email{yc3544@columbia.edu}
\affiliation{%
  \institution{Columbia University}
  \city{New York}
  \state{NY}
  \country{USA}
  \postcode{10027}
}

\begin{abstract}

Portfolio management is the art and science in fiance that concerns continuous reallocation of funds and assets across financial instruments to meet the desired returns to risk profile. Deep reinforcement learning (RL) has gained increasing interest in portfolio management, where RL agents are trained base on financial data to optimize the asset reallocation process. Though there are prior efforts in trying to combine RL and portfolio management, previous works did not consider practical aspects such as transaction costs or short selling restrictions, limiting their applicability. To address these limitations, we propose a general RL framework for asset management that enables continuous asset weights, short selling and making decisions with relevant features. We compare the performance of three different RL algorithms: Policy Gradient with Actor-Critic (PGAC), Proximal Policy Optimization (PPO), and Evolution Strategies (ES) and demonstrate their advantages in a simulated environment with transaction costs. Our work aims to provide more options for utilizing RL frameworks in real-life asset management scenarios and can benefit further research in financial applications. 
\end{abstract}


\maketitle
\section{Introduction}
Portfolio optimization remains one of the most challenging problems in finance, involving the continuous reallocation of funds across different financial products to maximize returns while managing risks \cite{FigueroaLpez2005ASS,b1,b2}. In recent years, there has been a growing interest in utilizing deep reinforcement learning (RL) in portfolio management, where RL agents are trained to trade and receive rewards based on changes in portfolio value. Previous works have focused on discrete single-asset trading \cite{b3} or multi-asset management \cite{b5,b6}. However, most of these works did not consider practical aspects such as transaction costs or short selling restrictions, limiting their real-world applicability. In our experiment, we designed an environment that simulates the real market including market assumptions such as transaction costs based on retrospective financial data. RL agent will be trained using with the underlying environment.

Inspired by a previous framework for trading cryptocurrencies using deep RL \cite{b6}, we propose several improvements:

\textbf{(1)} We compare the performance of three different RL algorithms: Policy Gradient Actor Critic (PGAC) \cite{b11}, Proximal Policy Optimization (PPO) \cite{ppo}, and Evolution Strategy (ES) \cite{es} whereas vanilla policy gradient was the RL algorithm used from previous works \cite{b6}. PGAC and PPO have been shown to achieve much stronger results in various domains than the vanilla policy gradient. In contrast to policy gradient based method, ES adopts a different approach in solving RL problems. Under ES framework, automatic differentiation is not need to optimize the policy network, and there is no need to maintain a separate value network. 



\textbf{(2)} Our work allows for the asset weight to be a continuous value and enables short selling as a viable option in asset management. In real-world asset management, it is impossible to trade asset weights using only a limited set of discrete options, and short selling is often an important part of the investment strategy. Previous works \cite{b6} failed to consider these requirements. However, our work is not intended to prove that continuous weights or short selling strategies are more advantageous in asset management. Rather, our goal is to provide more options for utilizing RL frameworks in real-life asset management scenarios.

\textbf{(3)} Our work allows additional features as input for both policy network and state network. In previous work \cite{b6}, only the price from assets which are being traded is used as state input. However, in real-world asset management, it is unreasonable to make a decision without looking at extrinsic features in the financial market. We use a select set of features in our experiment but our framework is compatible with different type of features as input.

\section{Background}
\subsection{Portfolio optimization}

Dynamic portfolio optimization is the process of allocating optimal assets at different times to achieve high return in total portfolio value while maintain a manageable risk. Traditional portfolio management methods can be broadly classified into four categories: "Follow-the-Winner," "Follow-the-Loser," "Pattern-Matching," and "Meta-Learning" \cite{b7} \cite{inbook}. Classical approaches employ dynamic programming and convex optimization, which require a discrete action space \cite{b7}. In the era of deep learning \cite{mlfinance}, there are models that focus on "Pattern-Matching" and "Meta-Learning." The "Pattern-Matching" algorithm aims to predict the market distribution in the next time period based on historical data, while the "Meta-Learning" method uses a combination of other strategies and metadata on financial assets.

In each portfolio, there consists $M$ assets where
$W = [w_{1},w_{2}...w_{i}...w_{n}]$ denote the N-dimensional weight vector subject to $\sum_{i=1}^{i=n}{w_{i}}=1$. Each $w_{i}$ is the weight allocated to asset $i$. 
$\xi = \left[\xi_{1}, \ldots, \xi_{n}\right]$ denotes the price vector where $\xi_{i}$ is the price of asset $i$. At time $t$, the total value of the portfolio $V_t = W_t \xi_t^T$. In the portfolio, we assume there is a risk-free asset that we can hold, denotes $w_r$ with fixed $\xi_r=1$. The existing machine learning approaches to portfolio optimization problems try to maximize $V_t - V_0$ as well as putting constraint on different risk measurements to avoid shortfall. 

\subsection{Reinforcement learning in portfolio optimization}

Reinforcement learning has been proposed as a suitable candidate for portfolio optimization by its own nature. According to \cite{Fischer2018ReinforcementLI}, reinforcement learning is able to combine both prediction and trading execution, which can be useful in high frequency trading. Moreover, the objective function of reinforcement learning can be designed to account for trading costs and other constraints, adding to its utility in this field.

In the portfolio optimization, the RL agent is trying to get positive reward in each time step $V_t-V_{t-1}$ by allocating weights $w_i$ for each asset in the portfolio. There are existing RL algorithms that manage to deal with this problem \cite{b5,b6}, however, there are several improvement can be made in their approaches. For instance, in the work of \cite{b5}, the authors did not consider transaction costs as a negative term in their reward function. The RL agent was not penalized when it's more actively trading. Therefore, the algorithm is not guaranteed to perform well in the realistic setting. Other works \cite{b6} take transaction costs in to account, yet they only consider long position for the assets in the portfolio. This may limit the performance of the RL agent by cutting of a powerful way of hedging.

The general reinforcement learning framework for portfolio management was proposed by \cite{b6} and used in \cite{b2}. They used different reinforcement learning algorithms to achieve promising results in different markets. In \cite{b6}, the authors introduced a general deterministic policy gradient algorithm to train the agent with different policy network architecture. The experiment was conducted in the crypto-currency market because its high volume and Bitcoin was its risk-free asset. 

\section{Environment Design and Data modeling}

 The RL environment generally contains three parts: State, Action, Reward. As the goal of our project is financial portfolio management, we set our action to be the portfolio weights corresponding to each asset for each time period. Assume we have n assets in our portfolio and our total portfolio weight is 1. For $t$'th trading period, we denote the action as a vector of portfolio weights: 
 $W_t = [w_{1,t},w_{2,t}...w_{i,t}...w_{n,t}]$ s.t. $\sum_{i=1}^{i=n}{w_{i,t}}=1$. 
 
 For each trading period, the agent submits $W_t$ to the environment(trading market) and get a reward $r_t$, which is the log return of our portfolio from time $t-1$ to $t$. Let $p_t$ denote the portfolio value at time $t$, then we define $r_t = \log{\frac{p_t}{p_{t-1}}}$. If trading cost is considered, for each trading period, $r_t$ needs to be modified to reflect the loss of trading cost. As suggested in Jiang's paper \cite{b6}, a shrinkage vector $\zeta$ can be applied to our portfolio. Assuming the cost of sell is $c_s$ and the cost of purchase is $c_p$, then $\zeta$ is recursively calculated in the following way suggested in Jiang's paper \cite{b6}.
 \begin{equation}
      \zeta_{t}=\frac{1}{1-c_{\mathrm{p}} w_{ 0,t}}
      \left[1-c_{\mathrm{p}} w_{ 0,t}^{\prime}-\left(c_{\mathrm{s}}+c_{\mathrm{p}}-c_{\mathrm{S}} c_{\mathrm{p}}\right) \sum_{i=1}^{m}\left(w_{i,t}^{\prime}-\zeta_{t} w_{i,t}\right)^{+}\right]
 \end{equation}
Denote the closing price for asset $i$ at period t to be $v_{i,t}$, and $V_t = [v_{1,t},v_{2,t}...v_{i,t}...v_{n,t}] $ then $w_{i,t}^{\prime} = \frac{v_{i,t}w_{i,t}}{W_{t-1} V_{t-1}^T} $  Hence, the reward after considering transaction cost becomes $R_t = \zeta r_t$ For each trading period, $R_t$ is returned to our agent as reward.

The state is constructed using a 3 dimensional tensor. For each time period t, we collect the most recent $d$ days of information for each feature and each asset. Those $d$ days will be referred as horizon. Features can be the open price, closing price, trading volume and etc. for corresponding asset. As the closing price is used to calculate the portfolio value, it must be included in the tensor. Let $e_{i,t,j}$ denote an entry in the tensor $E$. $e_{i,t,j}$ represents the information (or price or value) of asset $i$ at time $t$ for feature $j$. For $e \in E, i=1,2,...,n \quad  t =\tau, \tau-1, ...,\tau-d \quad  j = 1,2,...,m $, $m$ is the total number of features. At time t, the environment will return a tensor $E$ to our agent as a state. 

\section{Reinforcement Learning Algorithm}
\subsection{Policy Gradient Actor Critic}
In this paper, we use a policy gradient actor critic (PGAC) algorithm as our first Reinforcement Learning algorithm. Compared with the vanilla policy gradient algorithm used in the Jiang's paper \cite{b6}, PGAC algorithm is more robust and stable. Unlike vanilla policy gradient algorithm, PGAC has a separate neural network for the value function, which reduces its chance in converging into the local optimal\cite{Wen2021CharacterizingTG}. Meanwhile, by parameterizing the value in each state, PGAC helps the agent learn better and faster from the previous experience \cite{Wen2021CharacterizingTG}. 

The standard PGAC algorithm is described as the following\cite{DBLP:journals/corr/MnihBMGLHSK16}:\
\begin{algorithm}
\caption{Pseudocode for PGAC}
\SetAlgoNlRelativeSize{0}
\KwData{Policy $\pi_{\theta}(\mathbf{a} | \mathbf{s})$, Value Function $\hat{V}_{\phi}^{\pi}(\mathbf{s})$}
\KwResult{Updated Policy $\pi_{\theta}$}

\While{not converged}{
    \tcp{Sample data}
    \textbf{Step 1:} Sample $\left\{\mathbf{s}_{i}, \mathbf{a}_{i}\right\}$ from $\pi_{\theta}(\mathbf{a} | \mathbf{s})$\;
    
    \tcp{Fit value function}
    \textbf{Step 2:} Fit $\hat{V}_{\phi}^{\pi}(\mathbf{s})$ to sampled reward sums\;
    
    \tcp{Evaluate advantage}
    \textbf{Step 3:} $\hat{A}^{\pi}\left(\mathbf{s}_{i}, \mathbf{a}_{i}\right)=r\left(\mathbf{s}_{i}, \mathbf{a}_{i}\right)+\gamma \hat{V}_{\phi}^{\pi}\left(\mathbf{s}_{i}^{\prime}\right)-\hat{V}_{\phi}^{\pi}\left(\mathbf{s}_{i}\right)$\;
    
    \tcp{Update policy parameters}
    \textbf{Step 4:} $\nabla_{\theta} J(\theta) \approx \sum_{i} \nabla_{\theta} \log \pi_{\theta}\left(\mathbf{a}_{i} | \mathbf{s}_{i}\right) \hat{A}^{\pi}\left(\mathbf{s}_{i}, \mathbf{a}_{i}\right)$\;
    
    \textbf{Step 5:} $\theta \leftarrow \theta+\alpha \nabla_{\theta} J(\theta)$\;
}

\end{algorithm}

A few modifications were made to fit in portfolio management algorithm. First, time value of money is ignored as the training period is relative short. Therefore, we set discount rate $\gamma =1 $. Second, the action $a$ is a vector of portfolio weights, which should be continuous. This will affect the step 4 gradient calculation.

Rewrite step 4 to be 
\begin{equation}
\nabla_{\theta} J(\theta)=E_{\tau \sim \pi_{\theta}(\tau)}\left[\left(\sum_{t=1}^{T} \nabla_{\theta} \log \pi_{\theta}\left(\mathbf{a}_{t} | \mathbf{s}_{t}\right)\right)\left(\sum_{t=1}^{T} A\left(\mathbf{s}_{t}, \mathbf{a}_{t}\right)\right)\right]
\end{equation}

In the continuous space, to generate continuous actions, we use a normal distribution whose mean equals the output of the policy neural net. In this case, we add a Gaussian Noise into our system which helps it explore various states and actions. As we reach the later stage of training, we reduce the variance of the normal distribution to generate more accurate actions. Here is the policy formula:
\begin{equation}
\pi_{\theta}\left(\mathbf{a}_{t} | \mathbf{s}_{t}\right)=\mathcal{N}\left(f_{\text { neural network }}\left(\mathbf{s}_{t}\right) ; \Sigma\right)
\end{equation}

Another advantage of adding this Gaussian Noise is that it produces an explicit form of derivative, making it possible for neural networks to take gradient. As is shown below:

\begin{equation}
\log \pi_{\theta}\left(\mathbf{a}_{t} | \mathbf{s}_{t}\right)=-\frac{1}{2}\left\|f\left(\mathbf{s}_{t}\right)-\mathbf{a}_{t}\right\|_{\Sigma}^{2}+\mathrm{const}
\end{equation}

\begin{equation}
\nabla_{\theta} \log \pi_{\theta}\left(\mathbf{a}_{t} | \mathbf{s}_{t}\right)=-\frac{1}{2} \Sigma^{-1}\left(f\left(\mathbf{s}_{t}\right)-\mathbf{a}_{t}\right) \frac{d f}{d \theta}
\end{equation}

\subsection{Proximal Policy Optimization: PPO}

Proximal Policy Optimization (PPO) is a popular reinforcement learning algorithm that has gained attention for its ability to achieve state-of-the-art performance in a variety of tasks. Developed by OpenAI researchers in 2017, PPO builds upon the previous success of Trust Region Policy Optimization (TRPO)\cite{TRPO} by introducing an improved objective function that simplifies the optimization process. PPO uses a clipped surrogate objective function to ensure that the policy update steps are conservative, preventing large policy changes that could lead to instability. In this work, we follow the description in \cite{ppo} and experiment its performance in asset management.

\begin{equation}
L^{CLIP}(\theta) = \mathbb{E}_{t}\left[\min\left(r_t(\theta)\hat{A}_t, \text{clip}(r_t(\theta), 1-\epsilon, 1+\epsilon)\hat{A}_t\right)\right],
\end{equation}

where $r_t(\theta) = \frac{\pi_{\theta}(a_t|s_t)}{\pi_{\theta_{old}}(a_t|s_t)}$ is the probability ratio between the new policy and the old policy, $\hat{A}_t$ is the estimated advantage function, $\epsilon$ is a hyperparameter that controls the size of the trust region, and $\text{clip}(x,a,b)$ is a function that clips the value $x$ to be between $a$ and $b$. The objective of conservative policy iteration(CPI) is to maximize this function with respect to the policy parameters $\theta$, while ensuring that the policy update is conservative and does not deviate too far from the previous policy.

Following steps are followed to train PPO agent \cite{ppo}. 

\begin{algorithm}
\caption{Pseudocode for PPO}
\SetAlgoNlRelativeSize{0}
\KwData{Policy $\pi_{\theta}(\mathbf{a} | \mathbf{s})$, Value Function $\hat{V}_{\phi}^{\pi}(\mathbf{s})$}
\KwResult{Updated Policy $\pi_{\theta}$ and Value Function $\hat{V}_{\phi}^{\pi}$}

\While{not converged}{
    \textbf{Step 1:} Sample $\left\{\mathbf{s}_{i}, \mathbf{a}_{i}\right\}$ from $\pi_{\theta}(\mathbf{a} | \mathbf{s})$\;
    
    \textbf{Step 2:} Fit $\hat{V}_{\phi}^{\pi}(\mathbf{s})$ to sampled reward sums\;
    
    \textbf{Step 3:} Evaluate $\hat{A}^{\pi}\left(\mathbf{s}_{i}, \mathbf{a}_{i}\right)=r\left(\mathbf{s}_{i}, \mathbf{a}_{i}\right)+\gamma \hat{V}_{\phi}^{\pi}\left(\mathbf{s}_{i}^{\prime}\right)-\hat{V}_{\phi}^{\pi}\left(\mathbf{s}_{i}\right)$\;
    
    \textbf{Step 4:} Calculate the loss function for policy network\;
    $ J^{CLIP}(\theta) \approx \sum_{i} \min\left(r\left(\mathbf{s}_{i}, \mathbf{a}_{i}\right)\hat{A}^{\pi}\left(\mathbf{s}_{i}, \mathbf{a}_{i}\right), \text{clip}(r\left(\mathbf{s}_{i}, \mathbf{a}_{i}\right), 1-\epsilon, 1+\epsilon)\hat{A}^{\pi}\left(\mathbf{s}_{i}, \mathbf{a}_{i}\right)\right)$\;
    
    \textbf{Step 5:} Use gradient descent to update the policy network $\theta_{\text{policy}} \leftarrow \theta_{\text{policy}}+\alpha \nabla_{\theta_{\text{policy}}} J(\theta_{\text{policy}})$\;
    
    \textbf{Step 6:} Calculate the loss function for the value network\;
    $J^{VF}(\theta) \approx \sum_{i} \left(\hat{V}_{\phi}^{\pi}\left(\mathbf{s}_{i}\right) - V_{\phi}^{\pi}\left(\mathbf{s}_{i}\right)\right)^2$\;
    
    \textbf{Step 7:} Use gradient descent to update the value network $\theta_{\text{value}} \leftarrow \theta_{\text{value}}+\alpha \nabla_{\theta_{\text{value}}} J(\theta_{\text{value}})$\;
}

\end{algorithm}

The neural network structures we experimented below can be used in both PGAC and PPO framework simply changing the final loss function. Unlike the modification we made in PGAC, We do not need to add Gaussian Noise for the PPO architecture. 

\subsection{Evolution Strategy}
 Being a close variation of policy gradient algorithm, evolution strategy is a class of black box optimization technique that is first proposed by Rechenberg and Eigen in 1973 \cite{esold}. In each iteration, we generate a population of policies through ran, and the best-performing policies are selected to produce the next generation. The new population is generated by perturbing the parameters of the selected policies with some noise, and the process is repeated until convergence.

The advantage of using evolution strategy is that it does not require computing the gradient of the objective function, which can be difficult or infeasible in some cases. Additionally, evolution strategy is relatively simple and can be used with a wide variety of policy architectures. However, it can be computationally expensive since it requires maintaining a population of policies and evaluating their performance at each iteration. In portfolio optimization, evolution strategy has been shown to be effective in finding good policies and outperforms traditional optimization methods in certain scenarios \cite{Korczak2001EvolutionSI}.

\begin{equation}
\mathbb{E}_{\theta \sim p_{\psi}} F(\theta)
\end{equation}
where theta is the policy of the agent which in this case is just the parameter of the action neural net. Then F maps the sequence of action, portfolio weights of each asset that are being managed, the agent obey with respect to particular policy of the neural net to the finite horizon undiscounted reward, the percentage portfolio value in compare to what it has at time 0. 
The gradient of this objective function is given by the following equation.
\begin{equation}
\nabla_{\psi} \mathbb{E}_{\theta \sim p_{\psi}} F(\theta)=\mathbb{E}_{\theta \sim p_{\psi}}\left\{F(\theta) \nabla_{\psi} \log p_{\psi}(\theta)\right\}
\end{equation}
In evolution strategy, instead of assuming randomized policy, we perform a Gaussian smoothing, adding a Gaussian noise to the policy, and treat the policy itself as deterministic instead. Then our objective function becomes
\begin{equation}
\mathbb{E}_{\theta \sim p_{\psi}} F(\theta)=\mathbb{E}_{\epsilon \sim N(0, I)} F(\theta+\sigma \epsilon)
\end{equation}
To compute the gradient of this new objective function, one can evaluate it by the R.H.S of the following equation.
\begin{equation}
    \nabla_{\theta} \mathbb{E}_{\epsilon \sim N(0, I)} F(\theta+\sigma \epsilon)=\frac{1}{\sigma} \mathbb{E}_{\epsilon \sim N(0, I)}\{F(\theta+\sigma \epsilon) \epsilon\}
\end{equation}
Evolution strategy agent is trained with the following steps\cite{es}:

\begin{algorithm}
    \caption{Pseudocode for Evolution Strategy}
    \KwData{Policy $\pi_{\theta}(\mathbf{a} | \mathbf{s})$}
    \KwResult{Updated Policy $\pi_{\theta}$}
    \For{$\mathbf{t} = 0,1,2 \cdots$}{
    \textbf{Step 1:} Sample $\epsilon_{1}, \cdots, \epsilon_{n} \sim \mathcal{N} \left(0,\textit{I}\right)$;

    \textbf{Step 2:} Compute returns $F_{i} = F\left(\theta_{t} +\sigma \epsilon_{i} \right)$ for $i=0,1,2 \cdots n$;

    \textbf{Step 3:} Set $\theta_{t+1} = \theta_{t} + \alpha\frac{1}{\epsilon n}
    \sum_{i=1}^{n}F_{i}\epsilon_{i}$;
    }
\end{algorithm}
Then the gradient of the objective is evaluated directly using samples from the rewards from query the policy adding their respective Gaussian noise in the environment. In this way, evolution strategy manages to estimate the gradient of original objective function without actually taking the gradient to which is referred as gradient-free method. To calculate the reward of one particular noise added policy, one can simply query the environment using this policy and get a total rewards at the end of this roll out. Then policy can be updated using samples of this iteration and repeat the procedure until a desired reward is achieved. While the same deep learning structures in PGAC and PPO can be applied to the evolution strategy architecture, it would be computational expansive to have large and complex neural networks to optimize using ES. We will discuss the choice of neural networks for ES in the following section.

\section{Neural Network Construction}
\subsection{Data}
In this work, our training data consists of 4 financial instruments (one of them being a riskless asset), 6 financial instruments as features, and 7300 time steps (daily feature data from Year 2000 to Year 2020). The assets include the daily close price of Euro-Bobl Futures (FGBS), Euro-Schatz Futures (FGBM), and Euro-Bund Futures (FGBL). The features include the daily price of common commodities (gold, copper, and crude oil), and the daily price of 5-year and 10-year US treasury bond futures, and the SP500 index. While there are countless features that can be used to trade our assets, this set of features is suggested by domain experts who consider them the most relevant features in trading FGBS, FGBM, and FGBL. The data is selected based on interest from our asset managers in collaboration, but such selection does not compromise the generalization of our framework and experiment, as there is nothing special about the assets we selected. The data can be downloaded from a public financial database.

Therefore, the input tensor has dimensions of $(4, 50, 7)$. The first dimension represents the assets we are interested in trading in our portfolio, namely FGBS, FGBM, and FGBL, as well as a risk-free asset. However, as the price of the risk-free asset will be constant, it is sufficient to have only 3 risky assets in the input tensor. For each dimension, we add features from 6 relevant financial instruments (or indices), resulting in a total of 7 features. We are interested in using the previous 50 days' price data for each financial instrument, as it is a common metric used in trading. The final dimension of the input tensor is $(3, 50, 7)$.

As suggested in \cite{b6}, Convolutional Neural Networks (CNNs) demonstrate better performance compared to Recurrent Neural Networks. In this work, we experiment with both CNNs and Long Short-Term Memory (LSTM) networks, following the same Neural Network structure described in \cite{b6}.
\subsubsection{Convolutional Neural Network: CNN}

\begin{figure}[!htb] 
\centering
\includegraphics[width=6in]{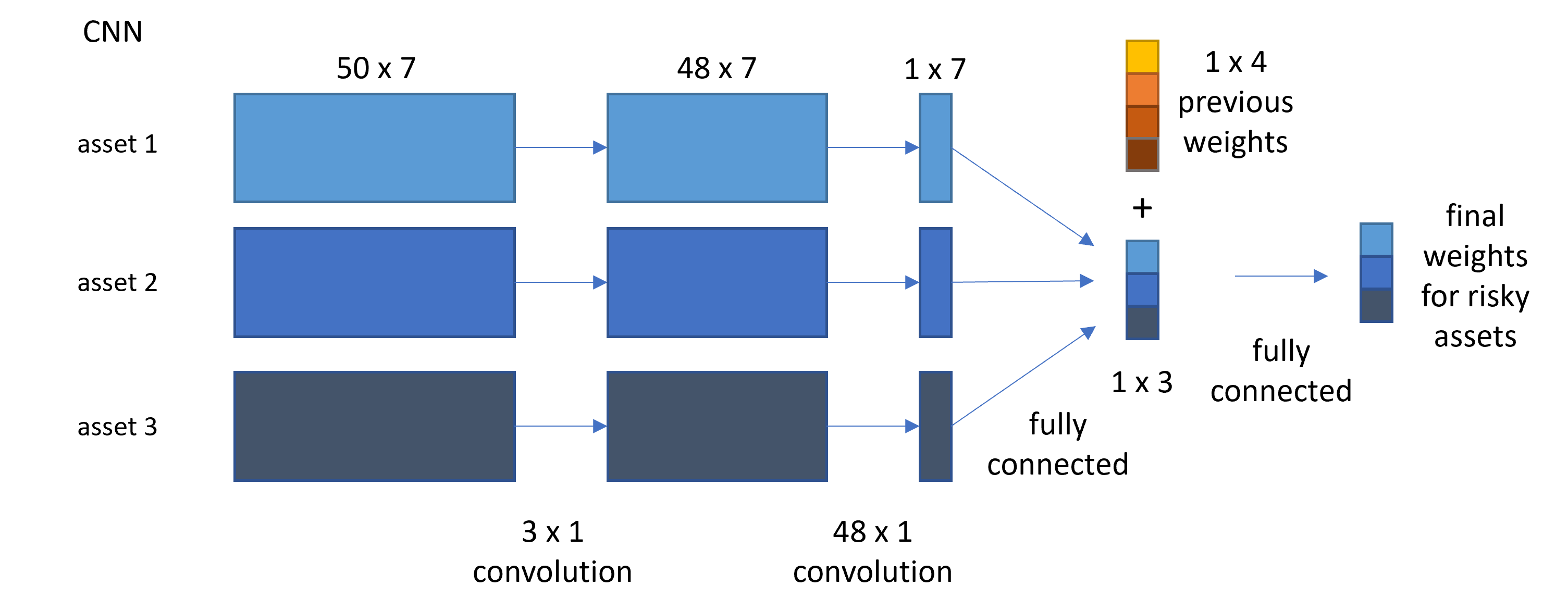}
\caption{Convolutional Neural Network: CNN}
\label{picture}
\end{figure}

To construct a CNN for the policy network, we followed the Ensemble of Identical Independent Evaluators (EIIE) structure described in \cite{b6}. The EIIE structure generates a representation for each asset in the current portfolio and uses a fully connected layer to aggregate all the representations and output the final weights. The previous portfolio weights are also added to keep track of the transaction cost when adjusting the portfolio weights.

The input tensor has a dimension of $n=3$, $m=7$, and $d=50$, where $n$ represents the number of risky assets in our portfolio, $m$ represents additional features that we consider, and $d$ represents the length of history for all the features. We first apply a convolution filter of size (1, 3) to extract signals from the price change within a window size of 3 days. We then get a tensor of size $n=4$, $m=7$, and $d-2=48$. Next, we apply a second convolution filter of size (1, 48) to output a vector of shape (3, 1). We concatenate this output vector with the previous actions (weights) of shape (4, 1), which gives us a feature map with size (7, 1). Finally, we add a fully connected layer and use "tanh" as the activation function. The output is a vector of length 3, where each value of this vector represents the mean of weights we should keep in the corresponding asset. The previous actions vector added to the neural network helps the agent learn the trading cost and avoid frequent trading. We use "tanh" as the last activation function to produce a continuous value range from -1 to 1, which not only enables shorting any asset but also keeps the weight of any asset within a reasonable range in accordance with real-life scenarios. In real trading scenarios, the total weights for any portfolio should sum up to 1, which can be achieved by setting the weight of the risk-less asset to be 1 minus the total weights for all risky assets.

A similar setting is used in the Value Net, except that no previous weights are added to the network, and the final output is a number representing the value of a state instead of a vector.

In training the RL algorithm, the daily price change tends to be very small, and the CNN may not be able to differentiate between very similar input tensors, leading to similar results. Following the idea from \cite{b6}, an amplification method is used before inputting the tensor into the neural network. Denote the original tensor as $E, e_{i,t,j}\in E$ and amplified tensor as $E',e_{i,t,j}' \in E'$. Set $e_{i,t,j}' = K \frac{e_{i,t,j}}{e_{i,t-1,j}} $ and when $t=1, e_{i,1,j}' = K$. In this work, K is set to be 100 which gives a reasonable scale on the input tensor.  

In the experiment, 50 days of data points is used for training and 10 days right after the training period is used for testing. 

\subsection{Recurrent Neural Network: LSTM}

\begin{figure}[!htb] 
\centering
\includegraphics[width=6in]{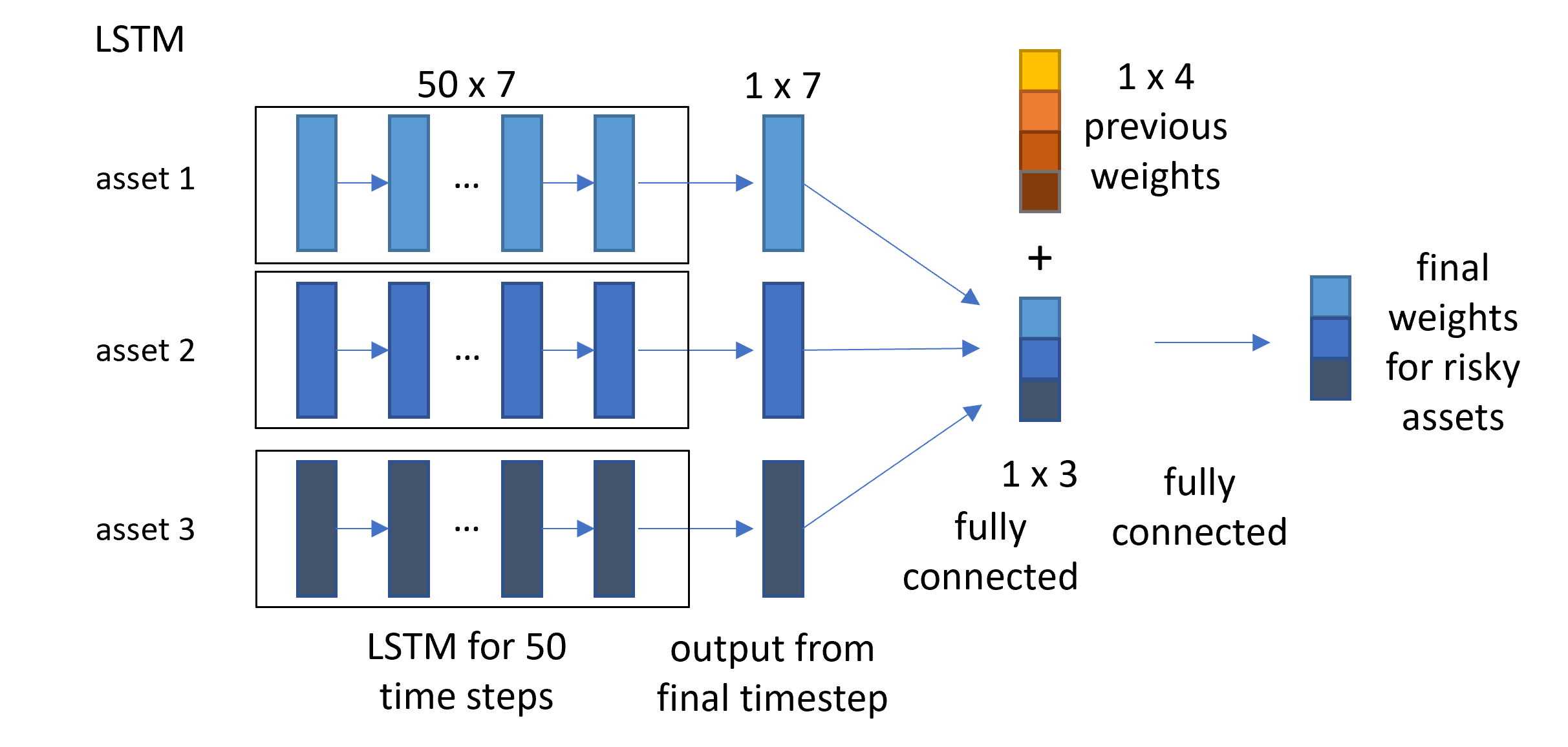}
\caption{Recurrent Neural Network: LSTM}
\label{picture}
\end{figure}
Recurrent Neural Networks (RNNs) are a type of neural networks that have shown remarkable success in modeling sequential data. Consequently, they have become a popular choice for the backbone of policy networks in reinforcement learning. The Long Short-Term Memory (LSTM) architecture, a variant of RNN, has gained wide recognition for its ability to capture long-term dependencies in time-series data. In our work, we leverage the power of LSTMs as the backbone for the policy network, enabling us to effectively model complex and dynamic environments.

To implement the LSTM structure, we follow the EIIE structure from \cite{b6}. Given the input tensor in the shape of $n = 3, m = 7, d = 50$, we initialize 3 independent LSTM networks. At each time step, the input is a vector of size 7 for each LSTM network, and we use 10 hidden units in each LSTM network. At the final time step, we concatenate the final output of LSTMs with a feedforward layer to generate the initial weight. Next, we append the previous portfolio weight to the initial weight vector, and a final feedforward layer is used to generate the final weight, with "tanh" as the activation function.

\subsection{Multilayer Perceptron for Evolution Strategy}

\begin{figure}[!htb] 
\centering
\includegraphics[width=6in]{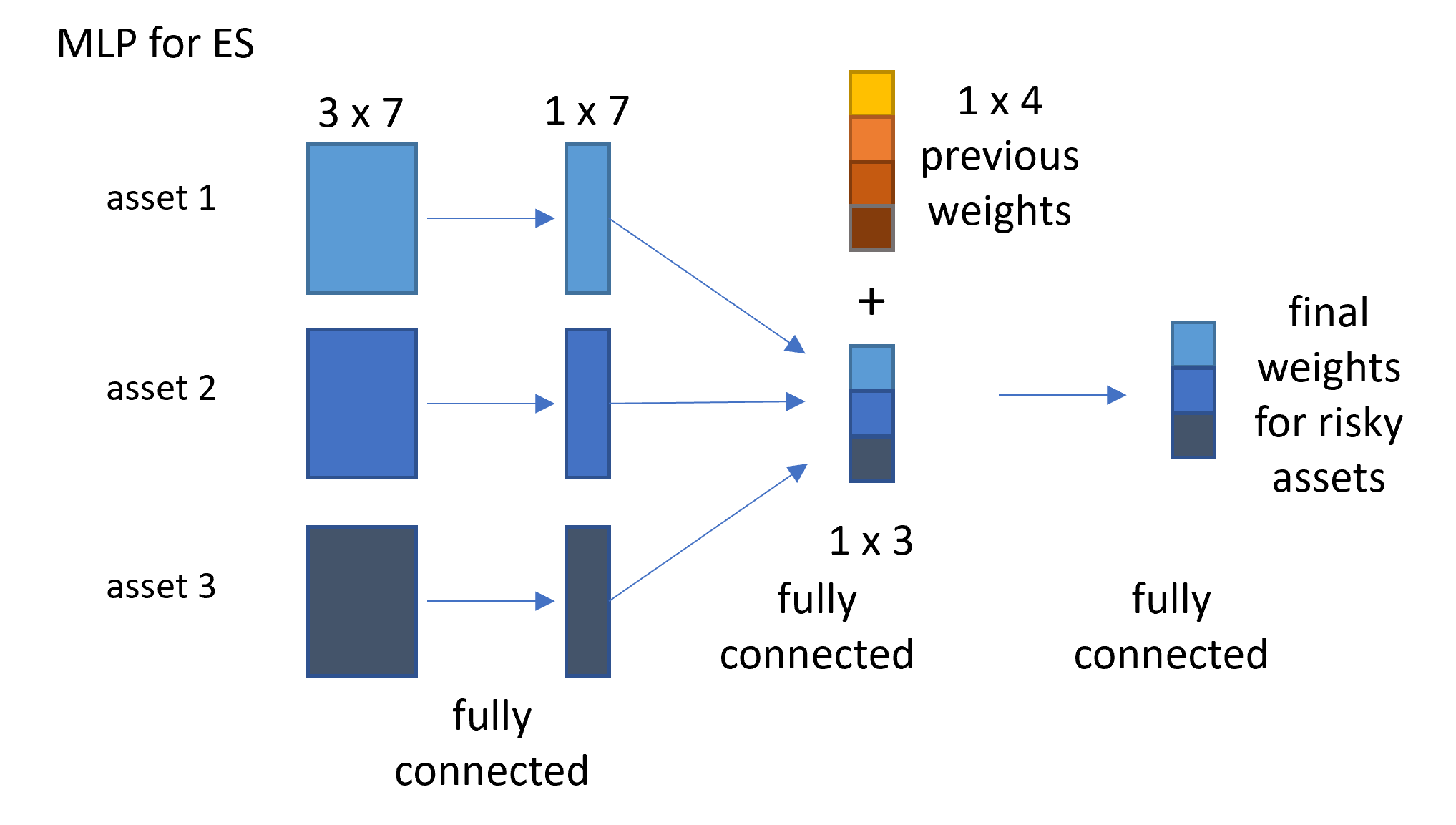}
\caption{Multilayer Perceptron for Evolution Strategy}
\label{picture}
\end{figure}

Since the state space that we designed for the portfolio management problem is a three-dimensional tensor, it is impractical and undesirable to use convolutional neural nets as the policy neural net. The curse of dimensionality implies that to obtain a good quality gradient estimate, an exponential amount of sample with respect to the policy space's dimensionality is required. Although there are techniques to address this issue, a large state space is generally unfavorable and can lead to an unstable policy during training. It is always wise to constrain the dimension of the state space. For this portfolio management problem, we truncate the horizon of the state tensor to 3, using three days of historical prices to make each portfolio weight decision. We choose standard fully connected layers in place of convolutional neural layers, which require us to flatten the input state space to one dimension. Then, our action net comprises two standard feedforward fully connected layers. We also leverage one of the significant advantages that the evolution strategy offers - ease of distribution. Additionally, we limit the maximum number of trades an agent can perform in a rollout to 50. It is conducive to restrict the maximum trading period during training to a small window since, in reality, trading history far back in the past has less importance than recent history. Finally, we test the trained policy for the next 10 days following the training period.

\section{Experiments}
\begin{table}
\captionsetup{position=bottom}
 \caption{Average Model Performance over 10 Trials}
  \centering
  \begin{tabular}{lllll}
    \toprule
    \cmidrule(r){1-2}
    Model     & Average Portfolio Value  & Average Sharpe Ratio & Average Maximum Drawdown \\
    \midrule      
PGAC -CNN        &    1.031  &     \textbf{1.20}   & 0.44\% \\
PGAC - LSTM      &    1.010  &     1.09   & 0.53\% \\
PPO - CNN        &    \textbf{1.044} &     1.17   & \textbf{0.30\%} \\
PPO - LSTM         &    0.988  &     -0.52  & 1.21\% \\
ES                 &    1.035  &     1.08   & 0.43\% \\
    \midrule  
Best Single Asset  &    1.025  &     1.13   & 0.69\% \\
Worst Single Asset &    0.982  &     -0.38  & 2.29\% \\
Follow the Winner  &    0.962  &     -0.73  & 0.97\% \\
Fixed Average Weight  &   1.02  &     1.15   & 0.57\% \\

    \bottomrule
  \end{tabular}
  \label{tab:table}
\end{table}
\subsection{Evaluation Settings}

To ensure a robust evaluation, we randomly sample 10 time spans from the 20 years of training data. Each sample consists of 50 days of training data, and we evaluate the RL algorithm on the following 10 days of data. For CNN and LSTM-based structures, which use a 50-day historical input tensor, we also use an additional 1 data points prior to the starting date to construct the tensor at time stamp 1. During evaluation, we freeze the policy network and value network. For each trial, we start with a portfolio value of 1 and calculate the final portfolio value, Sharpe Ratio, and Maximum Drawdown. We assume a trading cost of 0.01\% for each risky asset in our portfolio.

The formula for calculating Sharpe Ratio \cite{Sharpe49} is as follows:
\begin{equation}
\text{Sharpe Ratio} = \frac{P - 1}{\sigma_p}
\end{equation}
\\
Where $P$ is the final portfolio value, and $\sigma_p$ is the standard deviation of this portfolio over 10 days\\

The formula for calculating Maximum Drawdown is as follows:
\begin{equation}
\text{Maximum Drawdown} = \max_{i,j: j>i} \left(\frac{P_j - P_i}{P_i}\right)
\end{equation}
where $i, j$ represent the timestamp during evaluation: $ i, j \in [0, 10]$

We choose to evaluate the model's performance over a short time period because the financial signal that the model learns tends to decay over time. Our observations reveal that when we evaluate the model's performance beyond 10 days after the end of the training period, its performance drops immediately. Additionally, we avoid setting the training period too long because incorporating early information can introduce noise to the model. We have also observed that using 500 or 1000 days of data points actually impairs the model's performance.

\subsection{Results}
The results of the experiment for different RL settings are presented in Table 1. The CNN-based PPO model demonstrated the best performance in terms of portfolio value and maximum drawdown, while on average, the LSTM-based model did not perform as well as the CNN-based model. The ES-based model produced relatively good performance, considering the simplicity of the neural network used, and thus, it shows promising potential. It is worth noting that all RL-based strategies outperformed the rule-based strategies except for the LSTM-based PPO. This result is consistent with other RL works, where the PPO algorithm has demonstrated state-of-the-art performance. As noted in \cite{b6}, the LSTM network does not perform as well as the CNN under the EIIE structure, which is also observed in our experiment.

Although ES can be expensive to optimize, its benefit in financial applications is that it avoids the need to define the value of a state, which is a requirement in both PGAC and PPO. Moreover, the convergence assumption of actor-critic requires the agent to visit all states equivalently often, which can be a controversial assumption in financial applications.

In this experiment, we included a rule-based strategy called "Follow-the-Winner" as a baseline for comparison. This strategy simply purchases the asset that had the highest price increase in the last trading period. We also tested a fixed-average-weight strategy that holds the average weight for all the assets in our portfolio without making any trades in this period. This strategy has decent performance as it avoids all trading costs.

\subsection{Limitations}

Although we compared the average performance of RL algorithms based on 10 trials, we observed that RL algorithms can be unstable in learning signals across different trials due to the high variance of each RL agent. This is caused by the randomness from network training, value estimation, and optimization.

In this experiment, we did not compare any conventional machine learning algorithms with RL. This is because RL involves both signal discovery and execution, whereas conventional machine learning algorithms only generate trading signals. Therefore, it is not a fair nor valid comparison since the decision-making process in conventional machine learning algorithms is left to the portfolio manager.

It's important to note that we did not compare all possible network structures, hyperparameter settings, or RL algorithms. Additionally, we did not experiment with different financial data or features in our market. Our goal was not to identify a State-of-the-Art trading algorithm that would work in any market and in any time period, as the volatility and uncertainty in financial markets make this impossible. Instead, the goal of our work was to illustrate a potentially useful and generalized reinforcement learning framework in asset management, which can benefit further RL research in financial applications.

Although the RL agent conducts trades almost every trading period in our experiment, we recognize that this may not be realistic in asset management. We can impose additional trading restrictions, such as minimum trading frequency, maximum trading quantity, maximum asset weight, etc., based on each individual use case. Although these restrictions were not experimented with in our work, our framework is fully compatible with such constraints. For example, to impose a minimum trading frequency, we can simply take the agent's action every certain trading period. To impose a maximum trading quantity constraint, we can re-scale the final weight to be within a certain range for each asset from the policy network.

\section{Conclusion}

In summary, we propose a general framework for using RL algorithms (PGAC, PPO, ES) in asset management. Our framework allows for continuous asset weights, short selling, trade with extrinsic features. We tested our model using real-world trading data and demonstrated the advantages of RL algorithms in our limited experiment.

In reality, relying solely on reinforcement learning algorithms for trading or asset management can be extremely challenging. However, the power of RL - quick adaptation to dynamic environments and immediate execution based on perceived signals - can be valuable in financial applications.

\bibliographystyle{ACM-Reference-Format}
\bibliography{reference} 

\end{document}